\documentclass[conference]{IEEEtran}
\IEEEoverridecommandlockouts
\usepackage{cite}
\usepackage{amsmath,amssymb,amsfonts}
\usepackage{algorithmic}
\usepackage{graphicx}
\usepackage{textcomp}
\usepackage{xcolor}
\usepackage{hyperref}
\def\BibTeX{{\rm B\kern-.05em{\sc i\kern-.025em b}\kern-.08em
    T\kern-.1667em\lower.7ex\hbox{E}\kern-.125emX}}

\begin{document}

\title{Dynamic Encryption-Based Cloud Security Model using Facial Image and Password-based Key Generation for Multimedia Data}

\author{
\IEEEauthorblockN{Naima Sultana Ayesha\textsuperscript{1}, Mehrin Anannya\textsuperscript{1}, Md Biplob Hosen\textsuperscript{1,2}, Rashed Mazumder\textsuperscript{1}}
\IEEEauthorblockA{\textsuperscript{1}\textit{Institute of Information Technology}, \textit{Jahangirnagar University}, Savar, Dhaka, Bangladesh}
\IEEEauthorblockA{\textsuperscript{2}\textit{Department of Information Systems}, \textit{University of Maryland Baltimore County}, Maryland, United States}
}

\maketitle

\begin{abstract}
In this cloud-dependent era, various security techniques, such as encryption, steganography, and hybrid approaches, have been utilized in cloud computing to enhance security, maintain enormous storage capacity, and provide ease of access. However, the absence of data-type-specific encryption and decryption procedures renders multimedia data vulnerable. To address this issue, this study presents a dynamic encryption-based security architecture that adapts encryption methods to any file type, using keys generated from facial images and passwords. Four diverse datasets are created, each with a consistent size of 2GB, containing varying combinations of image, audio (MP3 and MPEG), video, text, CSV, PPT, and PDF files, to implement the proposed methodology. AES is used to encrypt image data, AES-CTR is employed for audio or video data to meet real-time streaming needs, and Blowfish is used for other types of data. Performance analysis on all four datasets is conducted using AWS servers, where DATASET-1 demonstrates the best performance compared to the others.
\end{abstract}


\section{Introduction}
There are numerous methods to ensure the highest level of data security in the cloud, including encryption, steganography, authentication, the use of firewalls and antimalware software, data backup, and monitoring of data access, among others. The categorization of biometric authentication is discussed in \cite{b1}. By utilizing biometric data, which are inherently difficult to replicate or compromise, this method not only enhances data security but also simplifies the key management process. Data integrity is maintained through cloud encryption techniques such as RSA, AES, and steganography \cite{b2,b3}. Digital images are the most commonly used cover media for steganography, with various methods depending on the embedding domain, size, and type of retrieval \cite{b4,b5}. The effectiveness of hybrid techniques that combine steganography and encryption is demonstrated in \cite{b6,b7,b8}. Text-based approaches are inadequate for large files, which presents challenges for multimedia encryption \cite{b9}. A study \cite{b10} examines the drawbacks and offers recommendations for future work, such as enhanced key generation, blockchain utilization, personalized encryption, and multispectral data encryption.\\
In cloud security, the use of biometric features for key generation and effective key management significantly enhances system security. Utilizing multimedia data necessitates research to improve security by leveraging facial images and adaptive encryption methods. Addressing existing encryption issues is essential to provide robust security for multimedia content using facial traits and flexible encryption techniques.\\
The limitations and challenges identified in current research give rise to several issues in the fields of cloud security and multimedia encryption. Relying on a single biometric feature for key generation can undermine cloud security if additional factors are not incorporated. Furthermore, the absence of data-type-specific encryption techniques creates vulnerabilities for data stored in the cloud.\\
Hybrid approaches introduce complexity, which can negatively impact key management, system performance, and data capacity. The proposed work aims to achieve robust key generation by combining facial images and passwords, while developing flexible data-type-specific encryption and decryption methods to ensure maximum security for multimedia data. Additionally, diverse datasets containing various data types will be created to validate the proposed methodology. This paper introduces a novel method to establish a robust security framework for cloud data by incorporating key generation using facial images and passwords, along with flexible encryption techniques.\\
The rest of the paper is structured as follows: Section II presents a literature review on related work. Section III introduces the proposed system model, including the system architecture and a description of each stage. Section IV analyzes and documents the results of the proposed system. Finally, Section V provides the conclusion of the proposed model.

\section{Literature Review}
\label{chap:2}
In the field of cloud computing, data security has been a prominent area of research, with various approaches, such as steganography, encryption, authentication, and hybrid methods, being developed. A summary of different studies on data security reveals a range of methodologies and their limitations. Markandey et al.\cite{b11} introduced an approach focused on information ownership, which reduces input/output costs and allows clients to verify data integrity on untrusted servers. However, their study does not address future research on data classification or secure, trust-based cloud solutions. Shabbir et al.\cite{b12} proposed a health information security system using the Modular Encryption Standard (MES), based on layered security measures. However, their method does not accommodate image-oriented datasets and is limited to textual data encryption. Farid et al.\cite{b13} suggested a new identity management paradigm for IoT and cloud-based healthcare, utilizing multimodal encrypted biometric traits. Despite this, their approach does not fully address classical identity-based security issues, such as replay and man-in-the-middle attacks. Future research is needed to test the model's authentication in various scenarios.

Sannidhan et al.\cite{b14} proposed enhancing online transaction security using a pseudo-random key sequence based on facial features extracted with the SURF technique. The main limitation here is the reliance on a single feature extraction technique, which may limit the randomness of the key. Hossain et al.\cite{b15} combined AES encryption with fingerprint biometrics and one-time passwords for cloud data security. However, their method's reliance on fingerprints indicates a need for more diverse biometric solutions. Hosny et al.\cite{b10} conducted an in-depth analysis of multimedia encryption, identifying several shortcomings and suggesting improvements, such as better key generation and blockchain utilization. They pointed out issues like the complexity of DNA encryption, limited variation in cellular automata, and a lack of balance between security and complexity. Yasser et al.\cite{b16} explored new chaotic-based multimedia encryption techniques using 2D modification models and proposed future research directions, including the incorporation of additional chaotic maps and security layers.

Tajammul et al.\cite{b17} presented an algorithm that generates a key from the supplied data and encrypts it accordingly. Their future work aims to integrate data compression for more efficient cloud storage. Corpuz et al.\cite{b18} modified the Blowfish encryption algorithm to cater to various data formats. However, the processing time for Excel data remains a limitation. Khwailleh et al.\cite{b19} introduced a dynamic encryption technique that adjusts encryption keys based on the data type. Their approach is expected to improve with further research, especially in the areas of image classification and encryption for health platforms. Namasudra et al.\cite{b20} proposed a DNA-based encryption strategy that creates a strong 1024-bit key by combining user characteristics and passwords for large multimedia files. However, there is significant room for improvement in authentication procedures within cloud computing. Sajay et al.\cite{b21} suggested a hybrid technique combining homomorphic encryption and Blowfish encryption to address cloud security issues, anticipating future advancements in algorithms and secure storage solutions.

Finally, Adee et al.\cite{b6} proposed a four-step data security paradigm incorporating Rivest-Shamir-Adleman (RSA), Advanced Encryption Standard (AES), identity-based encryption, and Least Significant Bit (LSB) steganography. They also highlighted the need for additional research to improve the combination and security of multimedia data.

\section{System Methodology}
\begin{figure*}[thp!]
    \centering
    \centering
        \centerline{\includegraphics[height=6cm, width = 10cm]{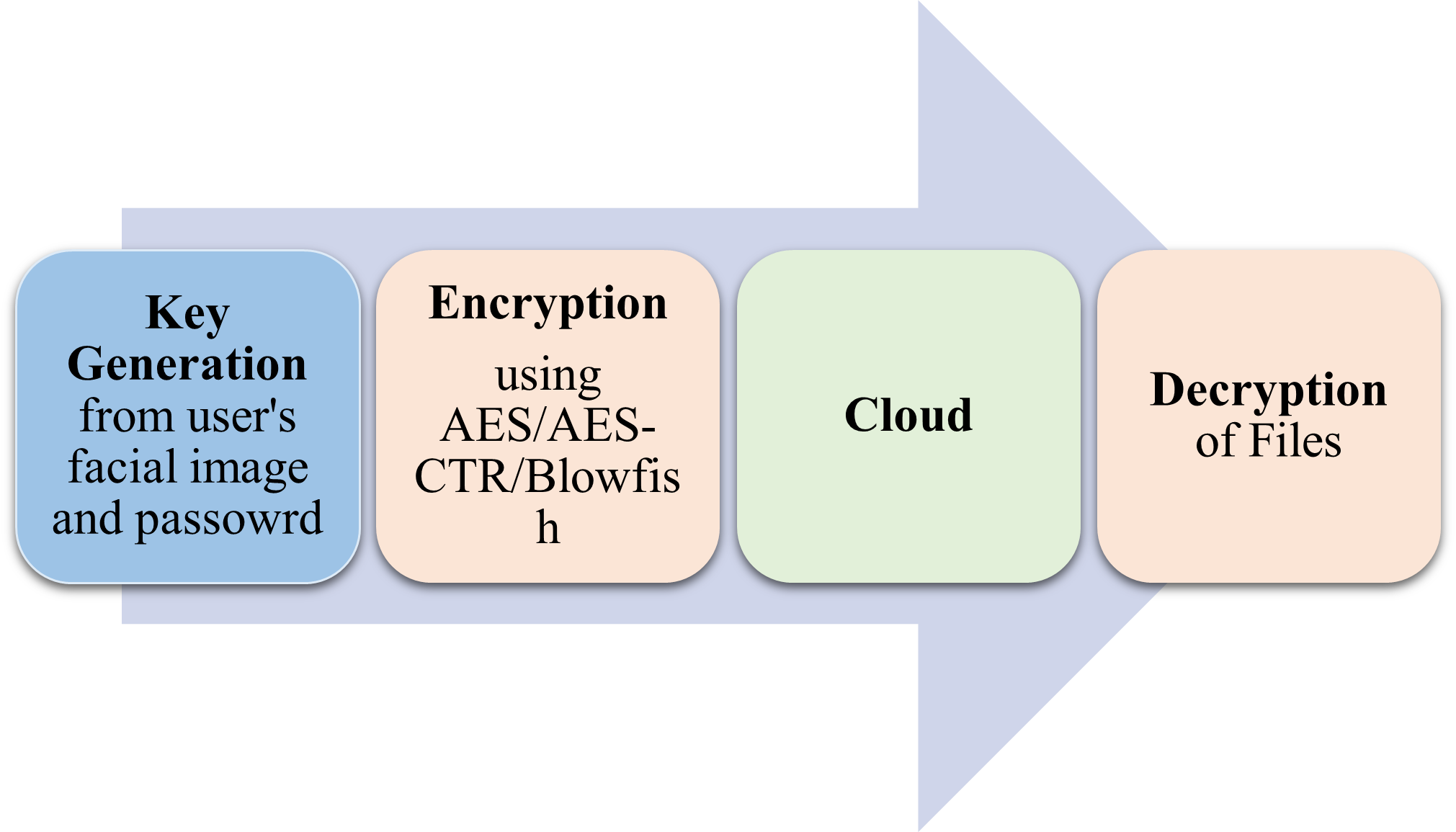}}
        \caption{Overview of system architecture.}
        \label{fig:motivation}
\end{figure*}

In this work, a cloud security model is proposed. Figure \ref{fig:motivation} illustrates the overall architecture of the system in brief. 
\begin{figure*}[thp!]
    \centering
    \centering
        \centerline{\includegraphics[height=6cm, width = 10cm]{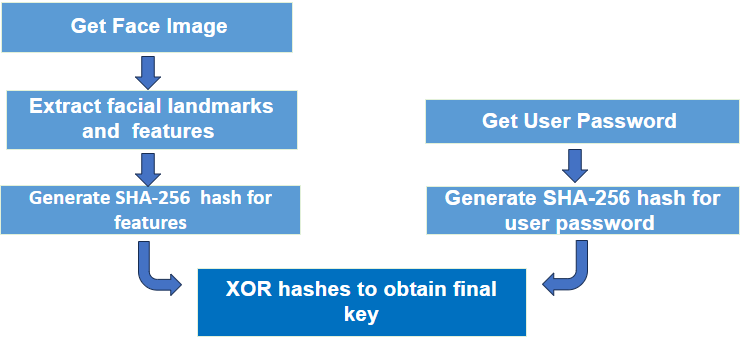}}
        \caption{Key Generation Process.}
        \label{fig:k}
\end{figure*}

\begin{figure*}[thp!]
    \centering
        \centerline{\includegraphics[height=6cm, width = 15cm]{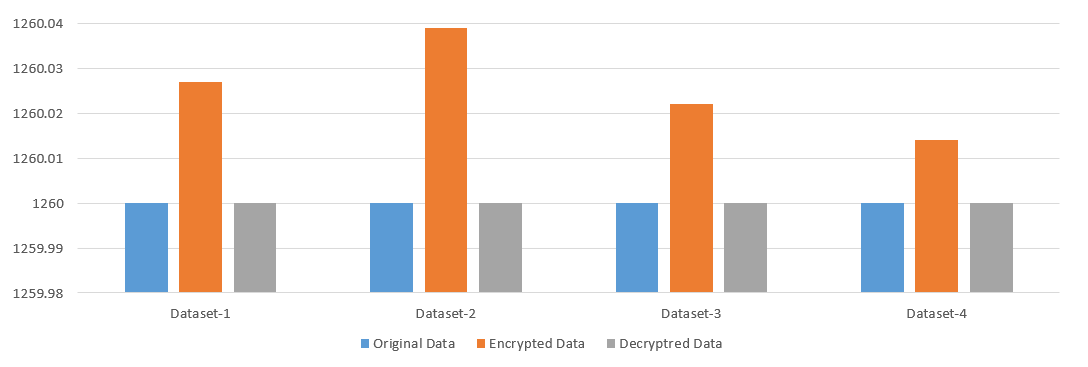}}
        \caption{Size variation among original, encrypted, and decrypted data.}
         \label{fig:mo}
\end{figure*}
In this system, distinctive facial features are extracted from face images, and passwords are provided by the user to create the keys needed for encrypting and decrypting data. Depending on the type of data being encrypted, the system integrates adaptive encryption techniques. AES-ECB is employed for image data, AES-CTR is used for audio and video data to support real-time streaming or random access playback, and flexible Blowfish encryption is applied for other data categories. Based on the required file name, data is retrieved and decrypted according to its type using the appropriate encryption key.

\subsection{Key Generation}
This proposed work generates keys by combining a password-based key with a key derived from facial features. This process is illustrated in Figure \ref{fig:k}. A pre-trained model, 'shape\_predictor\_68\_face\_landmarks.dat', is used to initialize the face detector and landmark predictor. The predictor first detects faces in the image and then identifies 68 distinct facial landmarks. The SHA-256 hash value of the encoded features, along with the given password, are generated. These two hash values are then combined bitwise using XOR, with their corresponding bytes, to produce a final key, which serves as the security key for both encryption and decryption.

\subsection{Encryption and Decryption Based on Data Type}
In this stage, different encryption-decryption methods are implemented based on the data type, all using the same generated key. For encrypting pixel block-structured image data, AES in ECB mode is effective, as it provides reliable encryption with easy decoding. During cloud uploads, padding ensures that the plaintext aligns with the AES block size. Similarly, AES in CTR mode efficiently encrypts audio data that is continuously transmitted, enabling fast playback and seeking. PKCS7 padding ensures the plaintext is properly sized for encryption, which can be decrypted with the provided AES key. With the same key and nonce used for encryption, AES-CTR’s stream cipher functionality supports efficient streaming for video content. Finally, the Blowfish algorithm strikes a balance between speed and security while offering flexibility in processing plaintexts of various lengths. It uses initialization vectors and padding to ensure the block size is appropriate, and during decryption, any padding applied is removed to yield the final, unpadded output.

\subsection{Dataset Description}
Four diverse datasets, each with a size of 1.2GB and containing different combinations of data types, have been created to implement the proposed methodology. These datasets are available in \cite{b22}. The system operates based on data type, and thus, various types of data, including image, audio, video, CSV, PPT, PDF, and text, were collected to create the customized datasets. The details of these datasets are provided in Table \ref{ds}.
\begin{table}[htbp]
\caption{Details of datasets}
\begin{center}
\begin{tabular}{|c|c|c|c|}
\hline
    \textbf{\begin{tabular}[c]{@{}l@{}}Dataset  \end{tabular}} & \textbf{\begin{tabular}[c]{@{}l@{}}Image (mb)  \end{tabular}} & \textbf{\begin{tabular}[c]{@{}l@{}}Audio-Video (mb) \\ \end{tabular}} & \textbf{\begin{tabular}[c]{@{}l@{}}Others(mb) \\ \end{tabular}} \\ \hline
    1 & 400  & 400 & 400 \\ \hline
    2 & 600 & 300 & 300 \\ \hline
    3 & 300 & 600 & 300 \\ \hline
    4 & 300 & 300 & 600 \\ \hline
\end{tabular}
\label{ds}
\end{center}
\end{table}

\section{Result and Discussion}

The proposed architecture was developed in an x64-based PC environment featuring an Intel(R) Core(TM) i5-6200U processor, 8GB of RAM, and running the Windows 11 operating system. The system uses Python 3.10.9 and Conda 23.3.1 for development. Using the generated key, four datasets were successfully encrypted and uploaded to a specific S3 bucket. Data retrieval and decryption are performed using corresponding keys and data-type-specific decryption methods.

As shown in Figure \ref{fig:mo}, the original and decrypted data sizes are identical, indicating no data loss during the encryption and decryption process. After encryption, the size of the data increased for all datasets, with Dataset-2 resulting in the largest increase in size.

Several evaluation measures, listed below, have been considered to provide a comprehensive performance study of the proposed model:

\subsection{Execution Time for Key Generation}

This measure evaluates the time (in seconds) taken by the proposed framework’s key generation operations, based on varying image sizes (5KB, 55KB, 150KB, 400KB, 850KB, 2MB, 5MB) and different password lengths (3, 6, 8, 10, 12). The time required to generate the key is presented in Table \ref{kgt}.

\begin{table*}[htbp]
\caption{Time measurement for  key generation}
\begin{center}
\begin{tabular}{|c|c|c|c|c|c|}
\hline
\textbf{Image Size} / \textbf{Password Length} & \textbf{3} & \textbf{6} & \textbf{8} & \textbf{10} & \textbf{12} \\ \hline
5KB & 0.01568 & 0.01598 & 0.01667 & 0.01693 & 0.01779  \\ \hline
55KB & 0.18755 & 0.21577 & 0.21776 & 0.23578 & 0.25242  \\ \hline
150KB & 0.34375 & 0.38859 & 0.61444 & 0.63400 & 0.65987   \\ \hline
400KB & 0.53120 & 0.67098 & 0.67676 & 0.701 & 0.70280                            \\ \hline
850KB & 0.61328 & 0.75904 & 0.77574 & 0.84460     & 0.93464                            \\ \hline
2.5MB & 2.19598 & 2.25511 & 2.24992 & 2.30883     & 2.31677                            \\ \hline
5MB & 3.33198 & 3.45086 & 3.54378 & 3.68518  & 3.76542\\        
\hline
\end{tabular}
\label{kgt}
\end{center}
\end{table*}

\begin{table*}[htbp]
\caption{Entropy Analysis For Key Generation}
\begin{center}
\begin{tabular}{|c|c|c|c|c|c|}
\hline
\textbf{Image Size} / \textbf{Password Length} & \textbf{3} & \textbf{6} & \textbf{8} & \textbf{10} & \textbf{12} \\
\hline
5KB   & 0.997 & 0.998 & 0.998 & 0.999 & 0.997 \\ \hline
55KB  & 0.997 & 0.996 & 0.992 & 0.999 & 0.999 \\ \hline
150KB & 0.999 & 0.999 & 0.999 & 0.999 & 0.994 \\ \hline
400KB & 0.998 & 0.999 & 0.999 & 0.998 & 0.999 \\ \hline
850KB & 0.998 & 0.994 & 0.976 & 0.995 & 0.999 \\ \hline
2.5MB & 0.997 & 0.997 & 0.994 & 1.00  & 1.000 \\ \hline
5MB   & 0.999 & 0.999 & 0.999 & 0.999 & 0.999 \\ \hline
\end{tabular}
\label{ne}
\end{center}
\end{table*}

\begin{table*}[htbp]
\caption{Time Taken for Brute Force Simulation}
\begin{center}
\begin{tabular}{|c|c|c|c|c|c|}
\hline
\textbf{Image Size} / \textbf{Password Length} & \textbf{3} & \textbf{6} & \textbf{8} & \textbf{10} & \textbf{12} \\
\hline
5KB   & 55.49 & 68.79 & 66.12 & 64.78  & 65.57 \\ \hline
55KB  & 57.74 & 68.75 & 62.18 & 65.98  & 67.83 \\ \hline
150KB & 60.28 & 89.40 & 90.95 & 89.33  & 91.61 \\ \hline
400KB & 60.97 & 86.80 & 91.77 & 92.30  & 98.25 \\ \hline
850KB & 66.36 & 89.57 & 90.30 & 90.49  & 92.76 \\ \hline
2.5MB & 63.44 & 66.17 & 64.88 & 66..99 & 67.89 \\ \hline
5MB   & 65.46 & 68.49 & 66.26 & 64.22  & 65.42 \\ \hline
\end{tabular}
\label{bfa}
\end{center}
\end{table*}

\begin{table*}[htbp]
\caption{Performance analysis on datasets}
\begin{center}
\begin{tabular}{|c|c|c|c|c|c|}
\hline
\textbf{Dataset} & \textbf{Total Datasize (mb)} & \textbf{Encryption time (sec)} & \textbf{Decryption time (sec)} & \textbf{Aggregate Throughput (mb/sec)} & \textbf{Size-overhead (mb)} \\
\hline
1 & 1260 & 10.94 & 10.02 & 115.19 & 0.027 \\ 
\hline
2 & 1260 & 8.89 & 8.46 & 141.70 & 0.039 \\
\hline
3 & 1260 & 8.45 & 7.53 & 148.97 & 0.022 \\
\hline
4 & 1260 & 12.15 & 12.25 & 103.63 & 0.014 \\
\hline
\end{tabular}
\label{per}
\end{center}
\end{table*}

It can be observed that for the same image size, the validation time for key generation remains minimal despite varying password lengths. However, as the image size increases, the key generation time also increases for the same password, and this variation becomes noticeable. Overall, the key generation time is minimal for smaller image sizes.

\subsection{Entropy Analysis}
The entropy is calculated based on the probabilities of '1' and '0' occurring in the binary data. The mathematical equations for this calculation are provided below:

\begin{equation}
\text{probability\_ones} = \frac{\text{ones\_count}}{\text{total\_bits}} \label{eq:p1}
\end{equation}

\begin{equation}
\text{probability\_zeroes} = \frac{\text{zeroes\_count}}{\text{total\_bits}} \label{eq:p0}
\end{equation}

\begin{align}
\text{entropy} &= - \left( \text{probability\_ones} \cdot \log_2(\text{probability\_ones}) \right. \notag \\
&\quad \left. + \text{probability\_zeroes} \cdot \log_2(\text{probability\_zeroes}) \right) \label{eq:p2}
\end{align}

\begin{equation}
\text{max\_entropy} = - \left( 0.5 \cdot \log_2(0.5) + 0.5 \cdot \log_2(0.5) \right) \label{eq:p3}
\end{equation}

\begin{equation}
\text{normalized\_entropy} = \frac{\text{entropy}}{\text{max\_entropy}} \label{eq:p4}
\end{equation}

Equations (\ref{eq:p1}) and (\ref{eq:p0}) calculate the probabilities of obtaining "one" and "zero" in a sequence of bits. Equation (\ref{eq:p2}) calculates the entropy, while equation (\ref{eq:p3}) computes the maximum entropy for the binary data of the key. Equation (\ref{eq:p4}) measures the normalized entropy by dividing the calculated entropy by the maximum entropy. The normalized entropy for the binary data of the key, as calculated in the assessment of the proposed framework, is shown in Table \ref{ne}, indicating the transition from unpredictability (1) to predictability (0). Since the entropy for key generation across different image and password size combinations is close to 1, it can be concluded that the key is highly unpredictable.

\subsection{Brute Force Attack Analysis}
A brute-force attack and simulation is a methodical, time-consuming process that tests every potential key in the key space to find the correct one. This approach is used to assess the proposed key-generating technique for different image sizes and password lengths. The time required for this simulation is shown in Table \ref{bfa}. The simulation takes varying amounts of time for different image sizes and password lengths, but it was unable to determine the password. This indicates that the key generation process is secure and that the generated key is strong enough to withstand such attacks.

\subsection{Encryption / Decryption Performance Analysis}
The performance analysis of the proposed method includes measurements of encryption and decryption time, aggregate throughput, size overhead, and other factors. The relevant mathematical equations are provided below:
\begin{equation}
\text{aggregate\_throughput} = \frac{\text{ data\_size}}{\text{ total\_time}} \label{e-at}
\end{equation}
    
 Equation (\ref{e-at}) calculates the aggregate throughput by dividing the total data size of the dataset by the total taken time.
\\
\begin{equation}
\text{size\_overhead} = \frac{\text{ length(crypted\_data)}}{\text{ data\_size}} \label{e-so}
\end{equation}
 Equation (\ref{e-so}) measures the size overhead by subtracting the original data size from the encrypted data size. The total size overhead for any dataset is calculated by summing the size overheads of each encrypted data item within the dataset. The performance study of the proposed method, applied to datasets with various data combinations, is presented in Table \ref{per}. The proposed method performs best on Dataset-3, which is enriched with audio and video data, followed by Dataset-2, which contains the most image data. Dataset-2 exhibits the highest size overhead, while Dataset-3 achieves the maximum aggregate throughput with the shortest encryption and decryption times. Dataset-1, with more balanced data sizes, provides better results overall. Dataset-4, which includes a broader variety of data types, shows longer processing times but lower size overhead and aggregate throughput.

\section{Conclusion}
This study proposes a dynamic encryption-based security architecture for cloud data, utilizing facial images and passwords for key generation. The system employs AES for image data, AES-CTR for audio/video data, and Blowfish for other data types, providing flexible encryption tailored to different content types. Performance analysis across four diverse datasets demonstrates that the proposed method enhances security for multimedia data, offering efficient encryption and decryption with minimal data loss. Despite the success in securing data, the method does face challenges such as potential accuracy issues from variations in image quality and environmental factors, privacy concerns regarding biometric data storage, and increased computational overhead from adaptive encryption.

To address these limitations, future research will explore biometric liveness detection beyond facial features and investigate privacy-preserving solutions for biometric data storage. Further work will also focus on optimizing encryption algorithms to reduce computational overhead without compromising security. Additionally, entropy analysis, brute-force simulations, and comparative studies using diverse datasets will be conducted to further validate and enhance the robustness of the proposed security model.

\bibliographystyle{ieeetr}
%\bibliographystyle{elsarticle-num}

%$\bibliographystyle{cas-model2-names}
\bibliography{ms1}

\begin{thebibliography}{10}

\bibitem{b1}
P.~Padma and S.~Srinivasan, ``A survey on biometric based authentication in cloud computing,'' in {\em 2016 International Conference on Inventive Computation Technologies (ICICT)}, vol.~1, pp.~1--5, 2016.

\bibitem{b2}
F.~F. Moghaddam, O.~Karimi, and M.~T. Alrashdan, ``A comparative study of applying real-time encryption in cloud computing environments,'' in {\em 2013 IEEE 2nd International Conference on Cloud Networking (CloudNet)}, pp.~185--189, IEEE, 2013.

\bibitem{b3}
S.~K. Singh, P.~Manjhi, and R.~Tiwari, ``Cloud computing security using steganography,'' {\em Journal of Emerging Technologies and Innovative Research,(JETIR)}, vol.~6, pp.~923--927, 2019.

\bibitem{b4}
A.~Y. AlKhamese, W.~R. Shabana, and I.~M. Hanafy, ``Data security in cloud computing using steganography: a review,'' in {\em 2019 International conference on innovative trends in computer engineering (ITCE)}, pp.~549--558, IEEE, 2019.

\bibitem{b5}
I.~J. Kadhim, P.~Premaratne, P.~J. Vial, and B.~Halloran, ``Comprehensive survey of image steganography: Techniques, evaluations, and trends in future research,'' {\em Neurocomputing}, vol.~335, pp.~299--326, 2019.

\bibitem{b6}
R.~Adee and H.~Mouratidis, ``A dynamic four-step data security model for data in cloud computing based on cryptography and steganography,'' {\em Sensors}, vol.~22, no.~3, p.~1109, 2022.

\bibitem{b7}
M.~S. Abbas, S.~S. Mahdi, and S.~A. Hussien, ``Security improvement of cloud data using hybrid cryptography and steganography,'' in {\em 2020 international conference on computer science and software engineering (CSASE)}, pp.~123--127, IEEE, 2020.

\bibitem{b8}
Z.~N. Al-Kateeb, M.~J. Al-Shamdeen, and F.~S. Al-Mukhtar, ``Encryption and steganography a secret data using circle shapes in colored images,'' in {\em Journal of Physics: Conference Series}, vol.~1591, p.~012019, IOP Publishing, 2020.

\bibitem{b9}
X.~Liu and A.~M. Eskicioglu, ``Selective encryption of multimedia content in distribution networks: Challenges and new directions,'' {\em IASTED Communications, Internet \& Information Technology (CIIT), USA}, p.~2003, 2003.

\bibitem{b10}
K.~M. Hosny, M.~A. Zaki, N.~A. Lashin, M.~M. Fouda, and H.~M. Hamza, ``Multimedia security using encryption: A survey,'' {\em IEEE Access}, vol.~11, pp.~63027--63056, 2023.

\bibitem{b11}
A.~Markandey, P.~Dhamdhere, and Y.~Gajmal, ``Data access security in cloud computing: A review,'' in {\em 2018 International Conference on Computing, Power and Communication Technologies (GUCON)}, pp.~633--636, IEEE, 2018.

\bibitem{b12}
M.~Shabbir, A.~Shabbir, C.~Iwendi, A.~R. Javed, M.~Rizwan, N.~Herencsar, and J.~C.-W. Lin, ``Enhancing security of health information using modular encryption standard in mobile cloud computing,'' {\em IEEE Access}, vol.~9, pp.~8820--8834, 2021.

\bibitem{b13}
F.~Farid, M.~Elkhodr, F.~Sabrina, F.~Ahamed, and E.~Gide, ``A smart biometric identity management framework for personalised iot and cloud computing-based healthcare services,'' {\em Sensors}, vol.~21, no.~2, p.~552, 2021.

\bibitem{b14}
M.~Sannidhan, K.~Sudeepa, J.~E. Martis, and A.~Bhandary, ``A novel key generation approach based on facial image features for stream cipher system,'' in {\em 2020 third international conference on smart systems and inventive technology (ICSSIT)}, pp.~956--962, IEEE, 2020.

\bibitem{b15}
M.~A. Hossain and M.~A. Al~Hasan, ``Improving cloud data security through hybrid verification technique based on biometrics and encryption system,'' {\em International Journal of Computers and Applications}, vol.~44, no.~5, pp.~455--464, 2022.

\bibitem{b16}
I.~Yasser, M.~A. Mohamed, A.~S. Samra, and F.~Khalifa, ``A chaotic-based encryption/decryption framework for secure multimedia communications,'' {\em Entropy}, vol.~22, no.~11, p.~1253, 2020.

\bibitem{b17}
M.~Tajammul and R.~Parveen, ``Auto encryption algorithm for uploading data on cloud storage,'' {\em International Journal of Information Technology}, vol.~12, no.~3, pp.~831--837, 2020.

\bibitem{b18}
R.~R. Corpuz, B.~D. Gerardo, and R.~P. Medina, ``Using a modified approach of blowfish algorithm for data security in cloud computing,'' in {\em Proceedings of the 6th International Conference on Information Technology: IoT and Smart City}, pp.~157--162, 2018.

\bibitem{b19}
D.~Khwailleh and F.~Al-Balas, ``A dynamic data encryption method based on addressing the data importance on the internet of things,'' {\em International Journal of Electrical and Computer Engineering (IJECE)}, vol.~12, no.~2, p.~2139, 2022.

\bibitem{b20}
S.~Namasudra, R.~Chakraborty, A.~Majumder, and N.~R. Moparthi, ``Securing multimedia by using dna-based encryption in the cloud computing environment,'' {\em ACM Transactions on Multimedia Computing, Communications, and Applications (TOMM)}, vol.~16, no.~3s, pp.~1--19, 2020.

\bibitem{b21}
K.~Sajay, S.~S. Babu, and Y.~Vijayalakshmi, ``Enhancing the security of cloud data using hybrid encryption algorithm,'' {\em Journal of Ambient Intelligence and Humanized Computing}, pp.~1--10, 2019.

\bibitem{b22}
{GitHub}, ``Dataset repository.'' \url{https://github.com/naisha275/Dataset}, 2024.
\newblock Accessed: 6 May 2024.

\end{thebibliography}

\end{document}